\documentclass{article}
\usepackage{emulateapj,apjfonts,psfig, epsfig}

\def\myputfigure#1#2#3#4#5%
{\hskip0.03\textwidth\vskip#5pt
\makebox[0pt]{\hskip#2in
\includegraphics[width=#3\textwidth]{#1}}\vskip#4pt\hfill}
\makeatletter

\newenvironment{tablehere}
  {\def\@captype{table}}
  {}
\makeatother

\def\fun#1#2{\lower3.6pt\vbox{\baselineskip0pt\lineskip.9pt
\ialign{$\mathsurround=0pt#1\hfil##\hfil$\crcr#2\crcr\sim\crcr}}}
\def\lap{\mathrel{\mathpalette\fun <}}

\def\mass{{\cal M}}

\def\msun{{\mass_\odot}}

\def\beq{\begin{equation}}
\def\eeq{\end{equation}}

\lefthead{Triaxial Nuclei}
\righthead{}

\begin{document}

\title{A Self-Consistent Study of Triaxial Black-Hole Nuclei}

\author{M. Y. Poon$^{1,2}$ and David Merritt$^2$}
\affil{$^1$Harvard-Smithsonian Center for Astrophysics, 60 Garden
Street, Cambridge, MA 02138;\\
$^2$Department of Physics and Astronomy, Rutgers University,
New Brunswick, NJ 08903}

\begin{abstract}
We construct models of triaxial galactic nuclei containing
central black holes using the method of orbital superposition, then
verify their stability by advancing $N$-body realizations of the models
forward in time.
We assume a power-law form for the stellar density,
$\rho \propto r^{-\gamma}$, with $\gamma=1$ and $\gamma=2$;
these values correspond approximately to
the nuclear density profiles of bright and faint galaxies respectively.
Equidensity surfaces are ellipsoids with fixed axis ratios.
The central black hole is represented by a Newtonian point mass.
We consider three triaxial shapes for each value of $\gamma$:
almost prolate, almost oblate and maximally triaxial.
Two kinds of orbital solution are attempted for each mass model:
the first including only regular orbits, the second
including chaotic orbits as well. 
We find that stable configurations exist, for both values of
$\gamma$, in the maximally triaxial and nearly-oblate cases;
however steady-state solutions in the nearly-prolate geometry could
not be found.
A large fraction of the mass, of order 50\% or more, could be assigned
to the chaotic orbits without inducing evolution.
Our results demonstrate that triaxiality may persist even within the
sphere of influence of the central black hole, and
that chaotic orbits may constitute an important building block of
galactic nuclei.
\end{abstract}

\section{Introduction}

Schwarzschild (1979) demonstrated how to construct self-consistent
models of stellar systems in the absence of analytic expressions
for the orbital integrals.
His method consists of three steps.
i) Represent the stellar system by a smooth density law and divide it
into discrete cells;
ii) compute a library of orbits in the potential corresponding to the
 assumed density law, and record the time spent by each orbit in the
cells;
iii) find a linear combination of orbits that reproduces the cell
masses.
Using his method, Schwarzschild (1979, 1982) demonstrated
self-consistency of
triaxial mass models with and without figure rotation.
Most of the orbits in his solutions were regular, i.e. non-chaotic
(Merritt 1980).
Subsequently, Statler (1987) found a variety of self-consistent solutions
for the integrable, or ``perfect,'' triaxial mass models in which all 
orbits are regular.

Models like these with large, constant-density cores are now known to be poor
representations of elliptical galaxies, almost all
of which have high central densities (\cite{cra93}; Ferrarese et al. 1994).
Stellar densities rise toward the center approximately as power laws,
$\rho \propto r^{-\gamma}$. Fainter galaxies have steeper cusps,
$\gamma \approx 2$, while brighter galaxies have weaker
cusps, $0\lap\gamma\lap 1$, and exhibit an obvious break in the surface
brightness profile.
Following this discovery, Schwarzschild (1993) investigated
triaxial models with singular density profiles, $\rho\sim r^{-2}$,
and Merritt \& Fridman (1996) constructed self-consistent solutions
for triaxial galaxies with both weak and strong central cusps.
A significant portion of the phase space in these models was found
to be occupied by stochastic orbits.
Furthermore, triaxial self-consistency could sometimes only be achieved
by including some stochastic orbits.
Models containing stochastic orbits can represent bona-fide equilibria
as long as the stochastic orbits are represented as fully-mixed
ensembles (Merritt \& Fridman 1996; Merritt \& Valluri 1996).

In the last decade, evidence has grown that supermassive black
holes are generic components of galactic nuclei (Ho 1999).
There are roughly a dozen galaxies in which a compelling case
for the presence of a supermassive black hole
can be made based on the kinematics
of stars or gas (Merritt \& Ferrarese 2001),
as well as a number of active galactic nuclei
in which the kinematics of the broad emission line region
imply the existence of a supermassive black hole
(Peterson 2002).
Inferred masses range from $\sim 10^{6}\msun$ to $\sim 10^{9.5}\msun$
and correlate well with stellar velocity dispersions
(e.g. \cite{fer01}) and bulge luminosities (e.g. \cite{mcd02}).

The possibility of maintaining triaxiality within a galactic
nucleus containing a supermassive black hole remains a topic of interest.
Very close to the black hole, the gravitational force can be
considered a perturbation to the Kepler problem and the phase space
is essentially regular (\cite{mev99}; \cite{sas00}; \cite{pom01},
hereafter Paper I).
Farther from the black hole, the fraction of chaotic orbits
increases, up to a radius where the enclosed stellar
mass is a few times the black hole mass; beyond this radius
essentially all centrophilic orbits are chaotic (Paper I).
The tube orbits remain mostly regular since they avoid the destabilizing
center.
The persistence of regular orbits throughout the region where the
gravitational force from the black hole dominates, leaves
open the possibility of constructing self-consistent solutions.
Furthermore there is growing observational evidence for the existence
of bar-like distortions at the very centers of galaxies
(e. g. \cite{ers02}).

In Paper II (Poon \& Merritt 2002), we presented preliminary results
showing that self-consistent and stable triaxial equilibria could
be constructed for power-law nuclei with certain axis ratios.
In this paper, we present a more detailed investigation of triaxial
black-hole nuclei. We find that stationary solutions are possible
only for certain shapes; mass models that are too near to
prolate axisymmetry always evolve toward axisymmetry.

The properties of the mass models are presented in \S2.
Orbital solutions for various shapes and
density profiles are presented in \S3 and their stability is tested
by N-body simulation in \S4 and \S5.
Limits on the chaotic mass fraction are discussed in \S6.
\S7 discusses some implications for the nuclear dynamics of galaxies.

\section{Mass Model}
We model the stellar distribution by the density law:
\begin{eqnarray}
\rho_{\star} &=& \rho_{\circ} m^{-\gamma},\\
m^2 &=& \frac{x^2}{a^2}+\frac{y^2}{b^2}+\frac{z^2}{c^2},
\label{density}
\end{eqnarray}
The equidensity surfaces are concentric ellipsoids with fixed axis 
ratios $a : b : c$ and the radial profile is a power law with
index $-\gamma$. We define the outer surface by the ellipsoid
$m=m_{out}$ and measure the triaxiality via the index $T$, where:
\begin{eqnarray}
T \equiv \frac{a^2-b^2}{a^2-c^2}.
\end{eqnarray}
Oblate and prolate galaxies have $T = 0$ and $T = 1$ respectively.
$T=0.5$ corresponds to a ``maximally triaxial'' nucleus.

We consider two types of nuclei, the weak cusp ($\gamma=1$)
and the strong cusp ($\gamma=2$), which correspond roughly
to the density profiles observed at the centers of bright and faint
elliptical galaxies respectively.
For each value of $\gamma$, we consider three shapes:
almost oblate ($T=0.25, a:b:c = 1.0:0.9:0.5$);
maximally triaxial ($T=0.50, a:b:c = 1.0:0.79:0.5$);
and almost prolate ($T=0.75$, $a:b:c=1.0:0.66:0.5$).
All models have $c/a=0.5$.

The black hole is represented by a central point mass with
$M_{bh} = 1$, which imposes a scale to the otherwise scale-free
stellar mass model.
We define two characteristic radii associated with the
presence of the black hole (Tables 1, 2).
i) $r_g$ is defined such that the enclosed stellar mass within
an ellipsoid with $m = r_g$ is equal to that of the black hole.
For $T=0.5$, $r_g = 0.64$ for $\gamma=1$ and $r_g=0.20$ for $\gamma=2$.
ii) $r_{ch}$ is the radius beyond which the regular, boxlike orbits
become almost all stochastic. For $T=0.50$, 
$r_{ch} \sim 2r_g$ for $\gamma=1$ and $r_{ch} \sim 6r_g$ for $\gamma=2$.

The gravitational potential can be obtained from Chandrasekhar's
theorem (\cite{chandra69}), which is valid for density laws
that are stratified on similar ellipsoids. 
The corresponding forces may be obtained in analytical form by
taking partial derivatives of the potential. The details are given in
\S 2 of Paper I.

Our orbital solutions require a finite outer radius.
Our aim was to explore the possibility of maintaining triaxiality
out to a radius of at least $r_{ch}$; hence
we chose the outer surface of our model to be large enough
that almost all of the density at $r_{ch}$ in a real galaxy
would be contributed by orbits with apocenters $r_+$ lying
below this surface.

To estimate $m_{out}$, we considered a spherical galaxy with density
\begin{eqnarray}
\rho_*^{sph} = \rho_{\circ} r^{-\gamma}.
\label{eq:rhor}
\end{eqnarray}
The isotropic distribution function corresponding to the
density law (\ref{eq:rhor})
can be found using Eddington's formula,
\begin{eqnarray}
\nonumber
f^{sph}(E)
     &=& \frac{\sqrt{2}}{4\pi^2}
\int_{E}^{a}\frac{d^2\rho_*^{sph}}{d\Phi^2}\frac{d\Phi}{\sqrt{\Phi-E}}
\\
     &+& \frac{\sqrt{2}}{4\pi^2}\lim_{{\Phi}\to\Phi_\infty
}\frac{d\rho_*^{sph}}{d\Phi}
        \frac{1}{\sqrt{\Phi-E}},
\end{eqnarray}
with $\Phi$ the potential generated by $\rho_*(r)$ and by the central
point
mass, and $\Phi_\infty \equiv \lim_{r\to \infty}\Phi(r)$.
We then change variables such that
\begin{eqnarray}
d\rho^{sph}_* =
\frac{2\pi}{r^2}\frac{f_\star^{sph}}{v_r}
                \Biggl| \frac{\partial(E, L^2)}{\partial(r_+, r_-)}
                \Biggr| dr_+dr_-,\label{drho}
\end{eqnarray}
with $v_r=\sqrt{2(E-L^2/2r^2-\Phi(r)}$ the radial velocity, and 
$r_+$ and $r_-$ the apocenter and pericenter of an orbit with
energy $E$ and angular momentum $L$.
The Jacobian is
\begin{eqnarray}
\frac{\partial(E, L^2)}{\partial(r_+, r_-)}
\equiv \left|\frac{\partial E}{\partial r_+}\frac{\partial L^2}{\partial
r_-}
     - \frac{\partial E}{\partial r_-}\frac{\partial L^2}{\partial
r_+}\right|.
\end{eqnarray}

According to (\ref{drho}), we define
\begin{eqnarray}
g(r_+, r) \equiv \int_{0}^{r_+}\frac{2\pi}{r^2}\frac{f_{\star}^{sph}}{v_r}
                \Biggl| \frac{\partial(E, L^2)}{\partial(r_+, r_-)}
                \Biggr |dr_-,\\
u(r_+, r) \equiv \frac{1}{\rho^{sph}_*(r)}\int_{0}^{r_+}g(r_+, r)dr_+.
\end{eqnarray}
$u(r_+, r)$ is the fraction of the mass at $r$ contributed by
orbits with apocenters $r \leq r_+ $.

\myputfigure{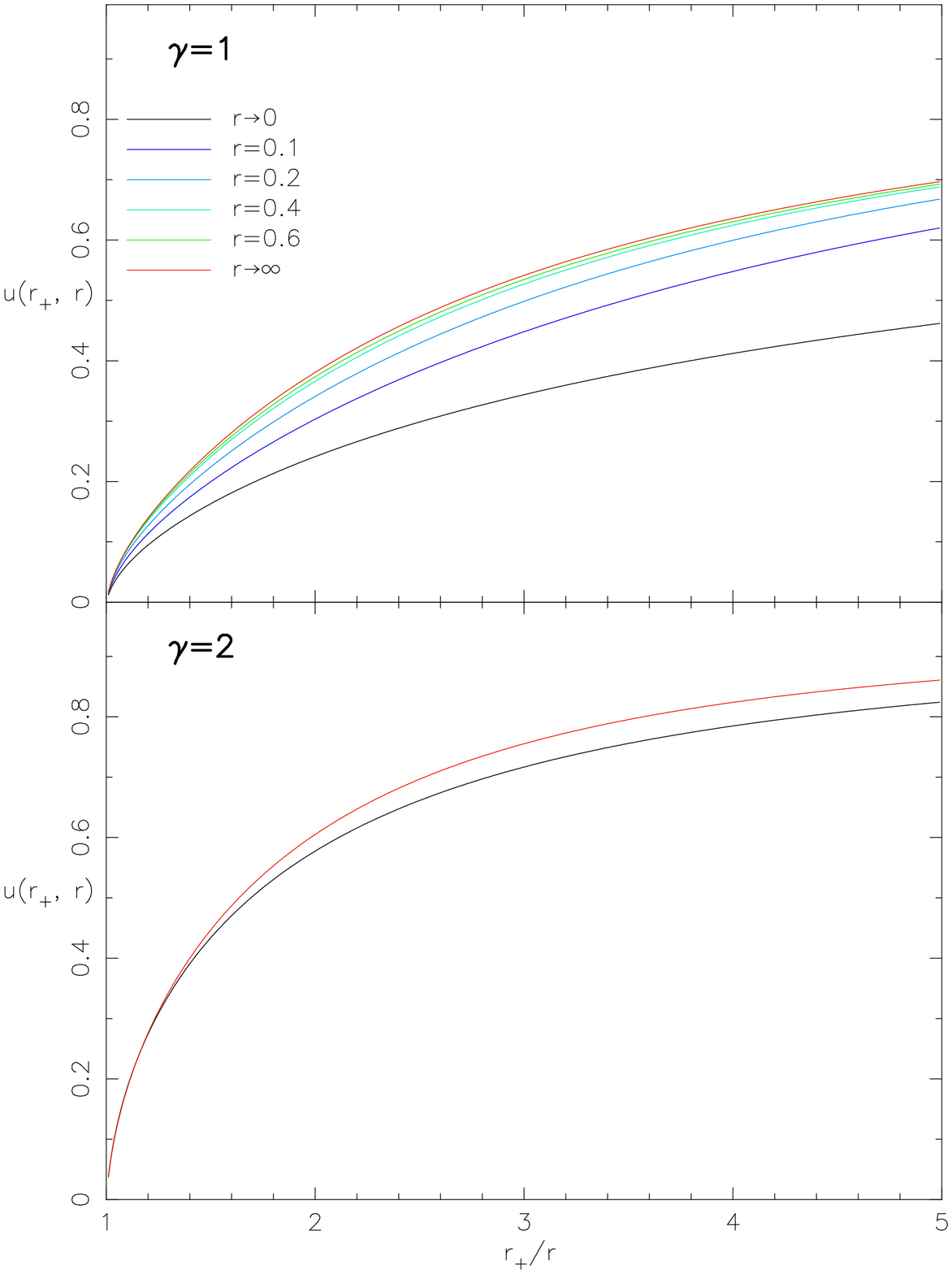}{3.2}{0.5}{-10}{-5}
\figcaption{\label{MYP_Figure1} $u(r_+, r)$ as a function of $r_+/r$ for
        $\gamma=1$ and $2$ and for different radii, as indicated.}
\vspace{\baselineskip}

Figure~\ref{MYP_Figure1} plots $u(r_+, r)$ as a function of $r_+/r$ for
both weak and strong cusp nuclei.
The red curves correspond to infinite $r$, i.e. no influence
from the black hole. The black curves correspond to very small $r$,
where the potential is almost Keplerian. $u(r_+, r)$ for the weak
cusp case is
smaller than that for the strong cusp at a given $r_+/r$,
because the density of the strong cusp falls off faster,
thus the mass at a given $r$ has to rely more heavily on nearby orbits.

Based on these results, we took $m_{out}$ to be $5r_{ch}$ for 
$\gamma=1$ and $3r_{ch}$ for $\gamma=2$, so that
about $70\% (75\%)$ of the mass at $r_{ch}$ is accounted for
in the weak (strong) cusp cases respectively. Tables~\ref{table1}
and \ref{table2} give the values of the characteristic radii in
model units.

\begin{tablehere}
\caption{Model Parameters for $\gamma=1$ \label{table1}}
\noindent\begin{tabular}{c|c|c|c}
$$ & $~~T=0.25~~$ & $~~T=0.50~~$ & $~~T=0.75~~$ \\ \hline 
$r_g$ & $0.594$ & $0.635$ & $0.694$ \\
$r_{ch}$ & $1.057$ & $1.128$ & $1.234$ \\
$m_{out}$ & $5.284$ & $5.642$ & $6.168$ \\ \hline
\end{tabular}
\end{tablehere}
\bigskip

\begin{tablehere}
\caption{Model Parameters for $\gamma=2$ \label{table2}}
\noindent\begin{tabular}{c|c|c|c}
$$ & $~~T=0.25~~$ & $~~T=0.50~~$ & $~~T=0.75~~$ \\ \hline 
$r_g$ & $0.177$ & $0.201$ & $0.241$ \\
$r_{ch}$ & $1.114$ & $1.270$ & $1.518$ \\
$m_{out}$ & $3.342$ & $3.811$ & $4.555$ \\ \hline
\end{tabular}
\end{tablehere}
\bigskip

\section{Construction of Orbital Solutions}

We followed standard procedures for constructing the Schwarzschild
solutions.
The cells used to define the mass distribution were defined as follows.
Each model with outer surface $m_{out}$ was divided into 64 shells.
The inner 63 shells were equidensity surfaces; the outermost
shell was an equipotential surface, in order to accomodate chaotic
orbits which fill regions defined by equipotential surfaces.
Shells were more closely spaced near the center.
The $42$nd shell in each model corresponded to $r_{ch}$.
Each shell was further divided into 48 angular cells per octant
as Merritt \& Fridman (1996), giving a total of 3072 cells per octant.
Due to the symmetry of the problem, it is necessary to consider only
a single octant when constructing the orbital solutions.

Orbits were computed in two initial condition spaces: stationary
start space, which yields mostly centrophilic orbits (pyramids,
stochastic orbits); 
and $X-Z$ start space, which yields mostly tube orbits.
Orbital energies were selected from a grid of 42 (52) values for
$\gamma=1$ (2), defined as the energies of equipotential surfaces
which were spaced similarly in radius to the equidensity shells.
The outermost energy shell, which is also an equipotential surface,
intersects the $x$-axis at $x=m_{out}$ for both mass models. 
$X-Z$ start space consists of orbits which begin on the $X$-$Z$ plane with 
$v_x=v_z=0$;
stationary start space consists of orbits which begin with zero velocity.
A total of 18144 (22464) orbits were integrated for 100 dynamical times
for $\gamma=1(2)$ and their contributions to the masses in the cells
were recorded.
In order to distinguish regular from stochastic trajectories,
the largest Liapounov exponent was computed for each orbit.

Twelve equations -- six representing the unperturbed motion and six
the linearized perturbations -- were integrated using the routine
RADAU of Hairer (1996), which is a variable time step, 
implicit Runge-Kutta Scheme which automatically switches between orders of 
5, 9, and 13. 
Energy was typically conserved to a few parts in $10^9$
over 100 orbital periods.

An orbit was considered regular if the largest Liapunov exponent
$\lambda$ satisfied $\lambda T_D< 10^{-0.9}$. 
The dynamical time $T_D$ is defined as the period of a circular
orbit of the same energy in the equivalent spherical potential,
which is defined to have a scale length $(abc)^{\frac{1}{3}}$
(Paper I).
This threshold was determined empirically by making 
histograms of Liapounov exponents of mono-energetic orbits at
various energies;
$\lambda T_D\approx 10^{-0.9}$ was found to always
separate the two peaks in the histogram
corresponding to regular and chaotic orbits.
A large fraction of the computed orbits were found to be stochastic,
as shown in Table~\ref{table3}.

\begin{tablehere}
\caption{Fraction of regular orbits in the orbit libraries\label{table3}}
\noindent\begin{tabular}{c|c|c}
$$ & $~~~~~~~\gamma=1~~~~~~~$ & $~~~~~~~\gamma=2~~~~~~~$ \\ \hline 
$~~T=0.25~~$ & $0.58$ & $0.72$ \\ 
$~~T=0.50~~$ & $0.54$ & $0.67$ \\
$~~T=0.75~~$ & $0.49$ & $0.65$ \\ \hline
\end{tabular}
\end{tablehere}
\bigskip

We then found the linear combination of orbits which best reproduced
the cell masses. 
We did this by varying the orbital occupation numbers $C_i$ to 
minimize the quantity $\chi^2$, defined as:
\begin{eqnarray} 
\chi^2 = \sum_{l=1}^{N}(D_{l} - \sum_{i=1}^{M}B_{li}C_{i})^2. \label{rChi}
\end{eqnarray}
$B_{li}$ is the time spent by the $i$-th orbit in the $l$-th cell, 
$D_l$ is the mass of the $l$-th cell, and
$C_i$ is the occupation number of the $i$-th orbit.
The quadratic programming problem was solved using the NAG 
Fortran library routine E04NFC.
We measured the discrepancies of the orbital solutions by the parameter:
\begin{eqnarray}
\Delta^2 = \frac{1}{N}\sum_{l=1}^{N}(1 - \frac{1}{D_{l}}\sum_{i=1}^{M}B_{li}C_{i})^2,
\end{eqnarray}
the mean error in the cell masses.

We constructed two solutions for each mass model:
one using only regular orbits, and the other using 
both regular and chaotic orbits.
Schwarzschild (1993) was the first to include chaotic orbits
in self-consistent solutions.
Schwarzschild treated the chaotic orbits like regular orbits,
giving each chaotic orbit its own occupation number, even
though many of the chaotic orbits in his models were ``sticky''
and did not reach a time-averaged steady state during the interval of
integration.
He justified this practice by re-integrating all chaotic orbits
with non-zero occupation numbers for longer intervals in the fixed potential;
the extreme error in the cell masses
was found to increase from the original $\sim 1 \%$ to $\sim 10 \%$.
Merritt \& Fridman (1996) noted that fully-mixed chaotic orbits,
i.e. chaotic orbits that uniformly fill their accessible phase space
region, are bona-fide building blocks for steady-state galaxies and
approximated such building blocks by constructing ensemble averages
of the chaotic orbits at each energy.

We followed Schwarzschild in allowing different chaotic orbits
at a given energy to have different occupation numbers.
In our potentials, the chaotic orbits were observed to very quickly
fill the region accessible to them: their ``mixing times'' were
generally much shorter than the integration interval.
Hence we expected that our chaotic orbits would individually 
constitute bona fide building blocks for a stationary solution,
without the need to construct ensemble averages as in Merritt \& Fridman
(1996).
This expectation was confirmed by the $N$-body integrations 
described below.

\myputfigure{MYP_Figure2.ps}{3.2}{0.5}{-10}{-5}
\figcaption{\label{MYP_Figure2} 
Schematic diagram showing the three kinds of shells used for
constraining the orbital solutions.
(i) The innermost shells with all angular constraints imposed.
(ii) Intermediate shells with angular constraints ignored.
(iii) The outermost shells, which are ignored.
The highest-energy shell is an equipotential surface 
while the other shells are equidensity ellipsoids.}
\vspace{\baselineskip}

We could not find quadratic-programming
solutions which exactly reproduced the cell 
masses in all of the cells, including the outermost ones.
This is probably because the number of orbits visiting the outer shells 
is small, giving the quadratic programming algorithm little
freedom to fit the cell masses there.
Indeed in all of our self-consistent solutions,
most of the contribution to $\Delta^2$ 
came from the outermost cells.
We therefore considered solutions in which the
outermost cells were treated in various ways.
Figure~\ref{MYP_Figure2} illustrates the idea.
We defined three kinds of mass shells: 
(i) the innermost shells with all the constraints imposed, i.e. 
all of the angular cells included;
(ii) intermediate shells for which only the total mass was fit, 
with angular details ignored; 
(iii) the outermost shells which were excluded from the fit.
We carried out numerous tests in which we attempted to fit
various combinations of constraints. 
As expected, the discrepancy $\Delta$ defined by the innermost shells 
always decreased as the total number of constraints decreased. 
For instance, for the weak cusp model with $T=0.50$, 
including all the orbits allowed the innermost $60$ shells 
($r\sim 4 r_{ch}$) to be fit to machine precision when the 
outermost four shells were ignored.
When these ``exact'' solutions were advanced forward in time, however, 
they were found to exhibit significant evolution at large radii, 
due presumably to the poor fit in the outermost cells.
We discuss this further in \S 5.
Experiments like this persuaded us to focus on solutions that 
were not ``exact'' but rather were constrained to reproduce the 
densities in {\it all} cells within $m_{out}$ to as high an 
accuracy as possible. 
Such solutions exhibited smaller fractional errors in the inner 
cells ($r \lap r_{ch}$) than in the outer 
ones. Models constructed in this way were found to evolve much less than the 
``exact'' solutions and provide the basis for the discussion below.

\begin{figure*}[t]
\epsscale{1.5}
\plotone{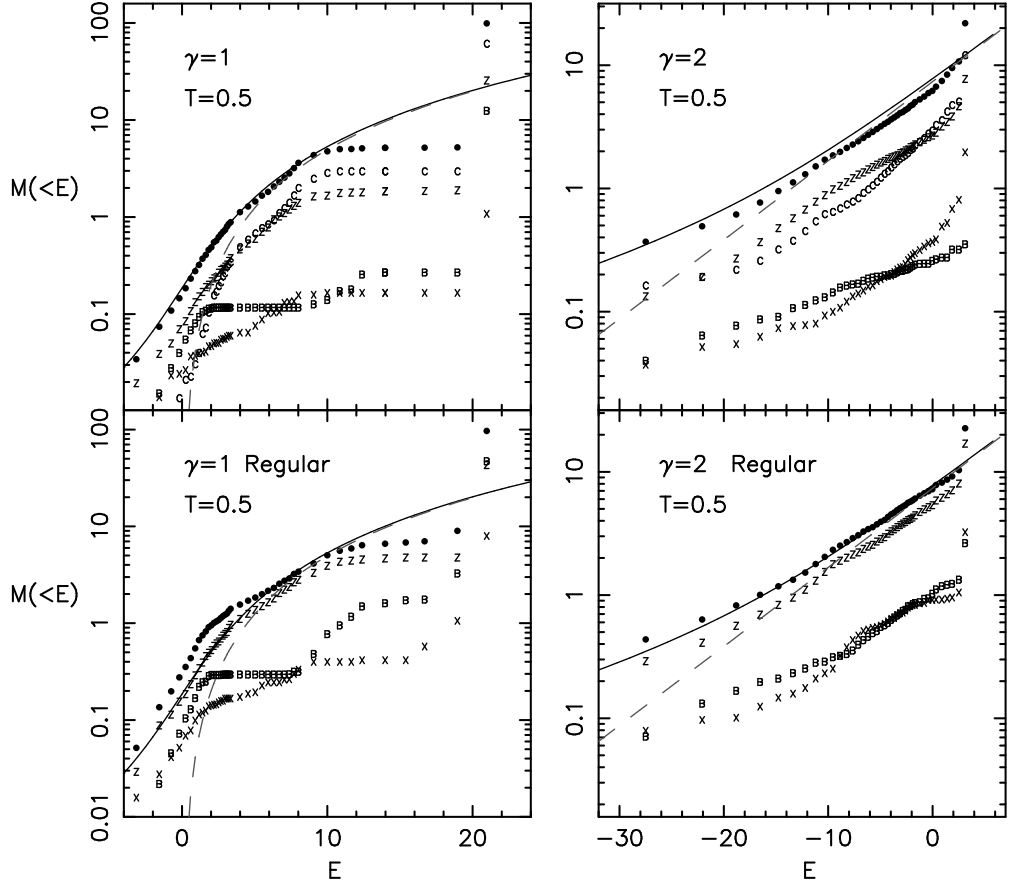}
\epsscale{1.5}
\caption{\label{MYP_Figure3}
Cumulative energy distributions of the various orbital families,
in self-consistent solutions with $T=0.5$.
The symbols ``B'', ``X'', ``Z'', ``C'' denote the 
mass contributed by box, $x$-tube, $z$-tube and chaotic orbits
respectively; black dots give the total.
Solid (dotted) lines show $M(<E)$ for the equivalent
spherical models defined in the Appendix, with (without) the central black hole.}
\end{figure*}

\begin{figure*}[t]
\epsscale{2.0}
\plotone{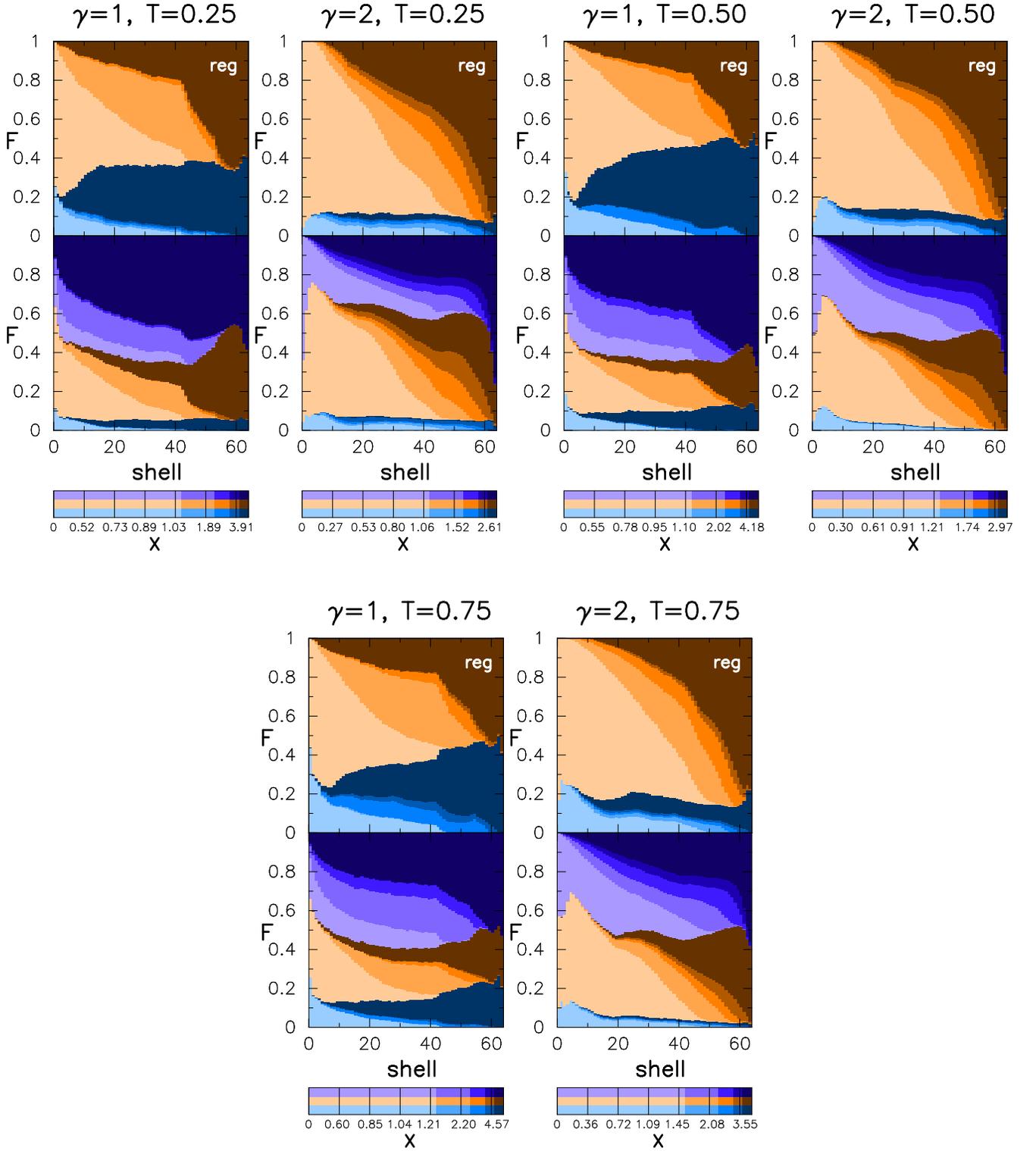}
\epsscale{1.0}
\caption{\label{MYP_Figure4}
Cumulative mass fraction $F$ contributed
by different kinds of orbits to different shells of the triaxial
solutions.
Box orbits are blue, tube orbits (both $z$- and $x$-tubes) are orange, 
and chaotic orbits are purple. 
Higher energies are represented by darker shades, defined according
to the $x$-intercept of the equipotential that has the same energy
as the orbit.
Numbers below the color bar indicate radii where the equidensity 
shells intersect the $x$-axis. 
Figures labelled ``reg'' respresent solutions constructed only using
regular orbits.}
\end{figure*}

The orbital content and the precision $\Delta$ of the solutions 
are represented in various ways in 
Tables~\ref{table4}-\ref{table6} and in
Figures~\ref{MYP_Figure3}-\ref{MYP_Figure4}.
We classify orbits into one of four families (cf. Paper I).
$x$- and $z$-tubes are regular orbits that circulate around the long
and short axes of the figure, respectively.
Pyramid orbits are the closest analogs to box orbits in these
potentials; they can be described as eccentric Keplerian
ellipses that precess due to torques from the stellar potential.
Their major elongation is contrary to that of the figure.
We will indiscriminately refer to these orbits as ``pyramids''
and ``box orbits'' in what follows.
The fourth family consists of the chaotic orbits.

Tables~\ref{table4} - \ref{table6} give the mass fractions
within various radii contributed by the four types of orbit.
$z$-tube orbits (short-axis tubes) are the biggest contributors in all of the 
regular-orbit solutions, making up at least $50\%$ of the total
mass and as much as 80\% in the nearly-oblate ($T=0.25$) models.
Pyramid orbits are next in importance; their contribution
reaches $\sim 40\%$ for $\gamma=1$ and $\sim 15\%$ for $\gamma=2$.
$x$-tubes (long-axis tubes) are relatively unimportant in the models
with $T=0.25$ and $0.5$, making up only a few percent of the total
mass.
In the nearly prolate models ($T=0.75$), their contribution
increases to $\sim 20\%$, however we argue below that these 
prolate solutions do not represent true equilibria.

When chaotic orbits are included in the orbit libraries,
the character of the solutions changes substantially:
at least $40\%$ of the mass is assigned to chaotic orbits
by the quadratic programming algorithm,
and as much as $60\%$ in the models with
$\gamma=1$ and $T=(0.25,0.5)$.
The inclusion of chaotic orbits lowers the mean error $\Delta$
by a modest amount in both the weak and strong cusp cases
(Tables 4-6). 
Much of the mass assigned to the chaotic orbits appears to be 
``taken'' from the pyramid orbits, as expected.
To our knowledge, these solutions contain a larger
fraction of chaotic orbits than in any other self-consistent
galaxy models.
Since the time-averaged shapes of the chaotic orbits are
nearly spherical, the existence of these solutions implies that
the regular orbits have more than enough freedom to reproduce the
triaxial shapes, and can do so even when forced to compensate
for the inauspicious shapes of the chaotic orbits.

Another way to represent the orbital makeup of the solutions 
is via the distribution of orbital energies.
Let $M_i(E)dE$ be the mass in orbits from the $i$th orbital family
whose energies lie in the range $E$ to $E+dE$.
The cumulative mass is given by $M_i(<E)=\int^E_{-\infty} M_i(E)dE$.
Figure~\ref{MYP_Figure3} shows $M_i(<E)$ for the self-consistent
solutions with $T=0.5$.
We show for comparison $M^{sph}(<E)$ computed for the equivalent 
spherical models defined in the Appendix.
One expects that $M^{sph}(<E)\approx\sum_i M_i(<E)$, and
Figure~\ref{MYP_Figure3} verifies that this is correct.
At high energies, there are discrepancies between the 
Schwarzschild solutions and the predictions of the spherical
models.
This is because the Schwarzschild solutions are truncated;
thus, orbits with the highest energies are forced to make 
large contributions,
which in turn lowers the contributions from orbits at lower energies.
This effect is more significant for the weak cusp models
which place most of their mass at large radii.
Nevertheless, at low energies even the $\gamma=1$ solutions
have nearly the same $M(<E)$ dependence as the equivalent spherical models.

\newpage

\begin{tablehere}
\caption{Orbital Content of Solutions with $T=0.25$ \label{table4}}
\noindent\begin{tabular}{ccccc}
\hline \hline
$$ & $z$-tubes & $x$-tubes & pyramids & chaotic\\ \hline 
$\gamma=1 :$ (regular)& &  &\\
$\log(\Delta) = -1.115$  & &  &\\
$r<0.5$ & 0.70 & 0.06 & 0.24 & --- \\ 
$r<1.0$ & 0.62 & 0.06 & 0.33 & ---\\
$r<1.5$ & 0.59 & 0.05 & 0.36 & ---\\ \hline
$\gamma=1 :$ (all)~~~~& &  &  &\\
$\log(\Delta) = -1.318$ & &  &  &\\
$r<0.5$ & 0.40 & 0.03 & 0.07 & 0.50\\ 
$r<1.0$ & 0.34 & 0.02 & 0.06 & 0.59\\
$r<1.5$ & 0.30 & 0.01 & 0.06 & 0.63\\ \hline

$\gamma=2 :$ (regular) & & & &\\
$\log(\Delta) = -1.437$& & & &\\
$r<0.4$ & 0.81 & 0.07 & 0.11 & --- \\
$r<0.8$ & 0.80 & 0.08 & 0.12 & --- \\
$r<1.2$ & 0.80 & 0.08 & 0.12 & --- \\ \hline
$\gamma=2 :$ (all)~~~~& &  &\\
$\log(\Delta) = -2.633$ &  & &\\
$r<0.4$ & 0.56& 0.03& 0.07& 0.34\\ 
$r<0.8$ & 0.55& 0.03& 0.07& 0.35\\
$r<1.2$ & 0.53& 0.03& 0.07& 0.37\\ \hline
\end{tabular}
\end{tablehere}

\newpage

\begin{tablehere}
\caption{Orbital Content of Solutions with $T=0.50$ \label{table5}}
\noindent\begin{tabular}{ccccc}
\hline \hline
$$ & $z$-tubes & $x$-tubes & pyramids & chaotic\\ \hline 
$\gamma=1 :$ (regular)& &  &\\
$\log(\Delta) = -1.169$& & &\\
$r<0.5$ & 0.64 & 0.13 & 0.23 & --- \\ 
$r<1.0$ & 0.56 & 0.08 & 0.36 & ---\\
$r<1.5$ & 0.52 & 0.07 & 0.41 & ---\\ \hline
$\gamma=1 :$ (all)~~~~&  &  &\\
$\log(\Delta) = -1.307$&  &  & &\\
$r<0.5$ & 0.29 & 0.07 & 0.11 & 0.54\\ 
$r<1.0$ & 0.27 & 0.03 & 0.10 & 0.60\\
$r<1.5$ & 0.25 & 0.02 & 0.10 & 0.62\\ \hline

$\gamma=2 :$ (regular) & & & &\\
$\log(\Delta) = -1.295$& & & &\\
$r<0.4$ & 0.73 & 0.11 & 0.16 & --- \\
$r<0.8$ & 0.73 & 0.11 & 0.15 & --- \\
$r<1.2$ & 0.73 & 0.12 & 0.15 & --- \\ \hline
$\gamma=2 :$ (all)~~~~& &  &\\ 
$\log(\Delta) = -2.567$&  & &\\ 
$r<0.4$ & 0.46& 0.05& 0.09& 0.40\\ 
$r<0.8$ & 0.45& 0.05& 0.06& 0.44\\
$r<1.2$ & 0.44& 0.05& 0.05& 0.46\\ \hline
\end{tabular}
\end{tablehere}

\begin{tablehere}
\caption{Orbital Content of Solutions with $T=0.75$ \label{table6}}
\noindent\begin{tabular}{ccccc}
\hline \hline
$$ & $z$-tubes & $x$-tubes & pyramids & chaotic\\ \hline 
$\gamma=1 :$ (regular)& &  &\\
$\log(\Delta) = -1.379$& & &\\
$r<0.5$ & 0.44 & 0.27 & 0.29 & --- \\ 
$r<1.0$ & 0.48 & 0.20 & 0.32 & ---\\
$r<1.5$ & 0.47 & 0.16 & 0.37 & ---\\ \hline
$\gamma=1 :$ (all)~~~~& &  &  &\\
$\log(\Delta) = -1.248$&  &  & &\\
$r<0.5$ & 0.27 & 0.14 & 0.16 & 0.42\\ 
$r<1.0$ & 0.26 & 0.07 & 0.14 & 0.53\\
$r<1.5$ & 0.22 & 0.06 & 0.16 & 0.56\\ \hline

$\gamma=2 :$ (regular) & & & &\\
$\log(\Delta) = -1.221$& & & &\\
$r<0.4$ & 0.57 & 0.20 & 0.23 & --- \\
$r<0.8$ & 0.59 & 0.21 & 0.20 & --- \\
$r<1.2$ & 0.58 & 0.22 & 0.20 & --- \\ \hline
$\gamma=2 :$ (all)~~~~& &  &\\ 
$\log(\Delta) = -2.471$ &  & &\\ 
$r<0.4$ & 0.38& 0.13& 0.11& 0.38\\ 
$r<0.8$ & 0.35& 0.12& 0.09& 0.44\\
$r<1.2$ & 0.33& 0.13& 0.08& 0.46\\ \hline
\end{tabular}
\end{tablehere}

Figure~\ref{MYP_Figure4} presents yet another representation of the
orbital populations of our self-consistent solutions.
We plot the cumulative mass fraction $F$ contributed by different
kinds of orbits to different shells of the model.
Box-orbits are blue, tube orbits are orange and chaotic orbits 
are purple; higher-energy orbits are represented by darker shades. 
Color bars relate the shades to the energy of the orbits. 
Numbers below the color bar indicate radii where the equidensity 
shells and equipotentials intersect the $x$-axis; 
note that the scale is not linear due to the higher resolution at 
lower radii.
For instance, in the weak cusp model with $T=0.50$, 
shell 20 is an equidensity ellipsoid with $x$-intercept $0.78$,
and regular box orbits with energy $E = \Phi(0.78, 0.0, 0.0)$ are 
represented by the lightest blue color.

At low energies, many of the $z$-tube orbits appear as saucers,
i.e. $2:1$ resonant orbits; at high energies, higher-order resonances
are important for these orbits.
It is surprising that $x$-tube orbits do not dominate the
nearly-prolate solutions with $T=0.75$.
However configuration-space plots of the $x$-tube orbits show that almost 
all of them are elongated contrary to the prolate figure: 
they are thin circular rings lying near the $y-z$ plane. 
Even in the nearly-prolate solutions, most of the mass is contributed
by $z$-tube orbits; the remaining contributions are mostly from 
high-energy box orbits, many of them associated with resonances.
In the nearly-prolate solutions, the most important resonances
are the ``fish'' ($2:3$) and the ``pretzels'' ($3:4$).
In the nearly-oblate solutions, the fish and the ``banana'' ($2:1$)
resonances dominate.
Many of the $z$-tubes and the high-energy boxes are replaced after the 
introduction of chaotic orbits (Figure~\ref{MYP_Figure4}). 

As discussed in more detail below, we were not able
to predict the long-term stability of a model based on its
value of $\Delta$.
Some models with fairly large $\Delta$'s were found to exhibit
almost no evolution, while other models with smaller $\Delta$'s
evolved significantly.
This suggests that the results of Schwarzschild modelling
should be interpreted with caution in cases where the long-term
stability of the solution has not been tested.

\section{$N$-body models}

Each of the solutions described above was found via a minimization
of the quantity $\chi^2$ describing the sum of the squared errors
in the cell masses (equation 10).
While the magnitude of $\chi^2$ might be expected to correlate with
the quality of the solution, there is really no way to know how
small $\chi^2$ must be in order for the solution to represent a
bona fide steady state, or how large a value of $\chi^2$ is consistent
with the existence of a smooth equilibrium solution.
One could require that each of the cell masses be fit exactly;
but we were never able to do this when the outermost shells were
included, and in any case, the discrete representation of the
density renders the interpretation of an ``exact'' solution problematic.
There are other uncertainties as well; for instance, both ``sticky'' chaotic
orbits and nearly-resonant, regular orbits may require very long integration
times before their cell masses approximate the steady-state values.

We tested whether our orbital superpositions represent true
equilibria by realizing them as $N$-body models and integrating
them forward in time in the gravitational potential computed
from the $N$ bodies themselves, and from a point mass representing
the black hole.
We prepared initial conditions corresponding to each of the orbital
solutions by re-integrating orbits with non-zero occupation numbers
and storing their positions and velocities at fixed time intervals. 
The sense of rotation of the tube orbits was chosen randomly.
We used $\sim 2 \times 10^6$ particles for representing the weak-cusp 
models and $\sim 2 \times 10^5$ particles for the strong-cusp models;
a smaller $N$ was chosen for $\gamma=2$ since the $N$-body integrations
were slower for the more condensed model.
We integrated the initial conditions forward in time using 
GADGET (\cite{syw01}), a parallel tree code with variable time 
steps.
The particle representing the black hole was allowed to move in 
response to the forces from the ``stars.'' 
The softening length was $0.005 (0.003)$ for the weak (strong) cusp cases. 
Energy was conserved to within $\sim 0.5 \%$ for the weak cusp models 
and $\sim 1 \%$  for the strong cusp models.

\begin{figure*}[t]
\epsscale{1.0}
\plotone{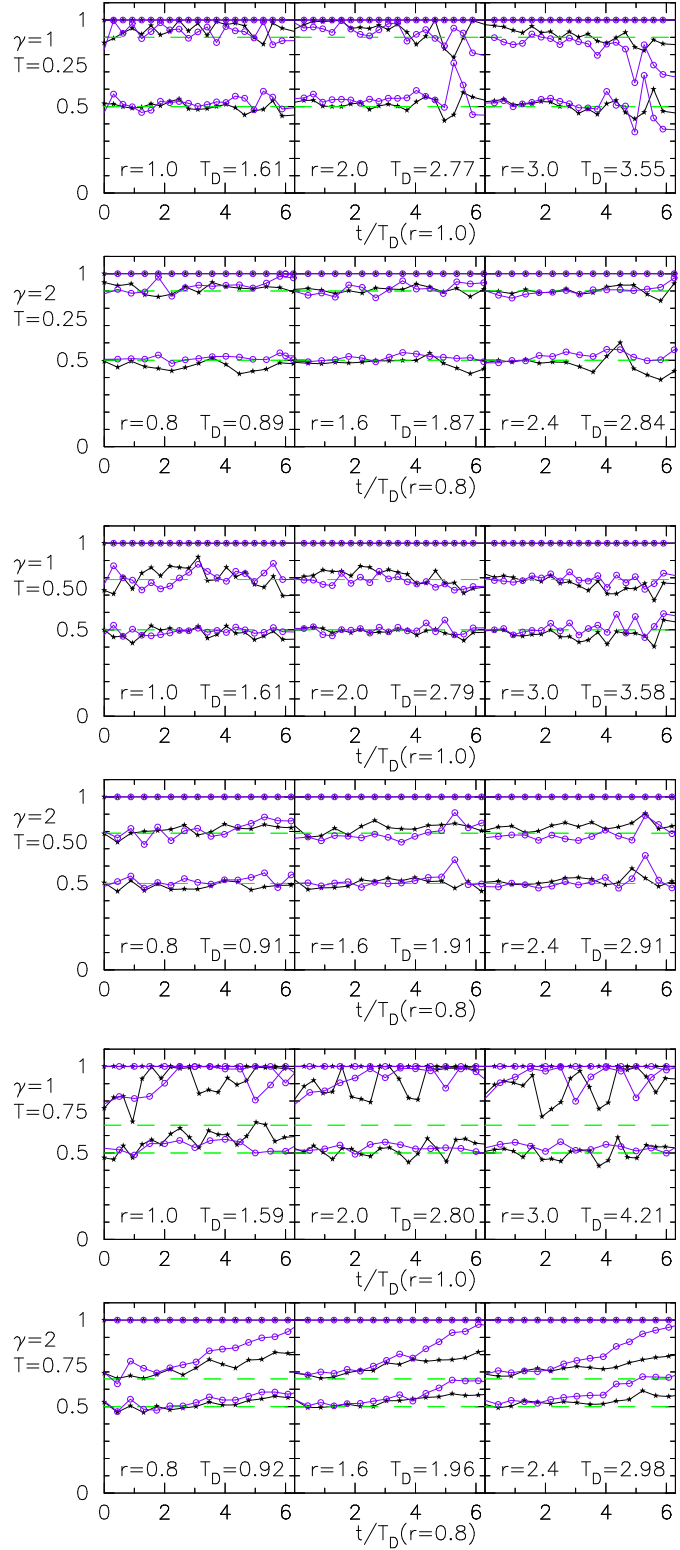}
\epsscale{1.0}
\caption{ Evolution of the axis lenghts at different radii for the 
orbital solutions shown in Figure~\ref{MYP_Figure4}.
The longest axis is defined to have unit length.
Black (purple) curves correspond to solutions containing 
only regular (both regular and chaotic) orbits.
Green lines indicate the values of the axis lengths in the
underlying mass models.
\label{MYP_Figure5}}
\end{figure*}
\bigskip

\begin{figure*}[t]
\epsscale{2.0}
\plotone{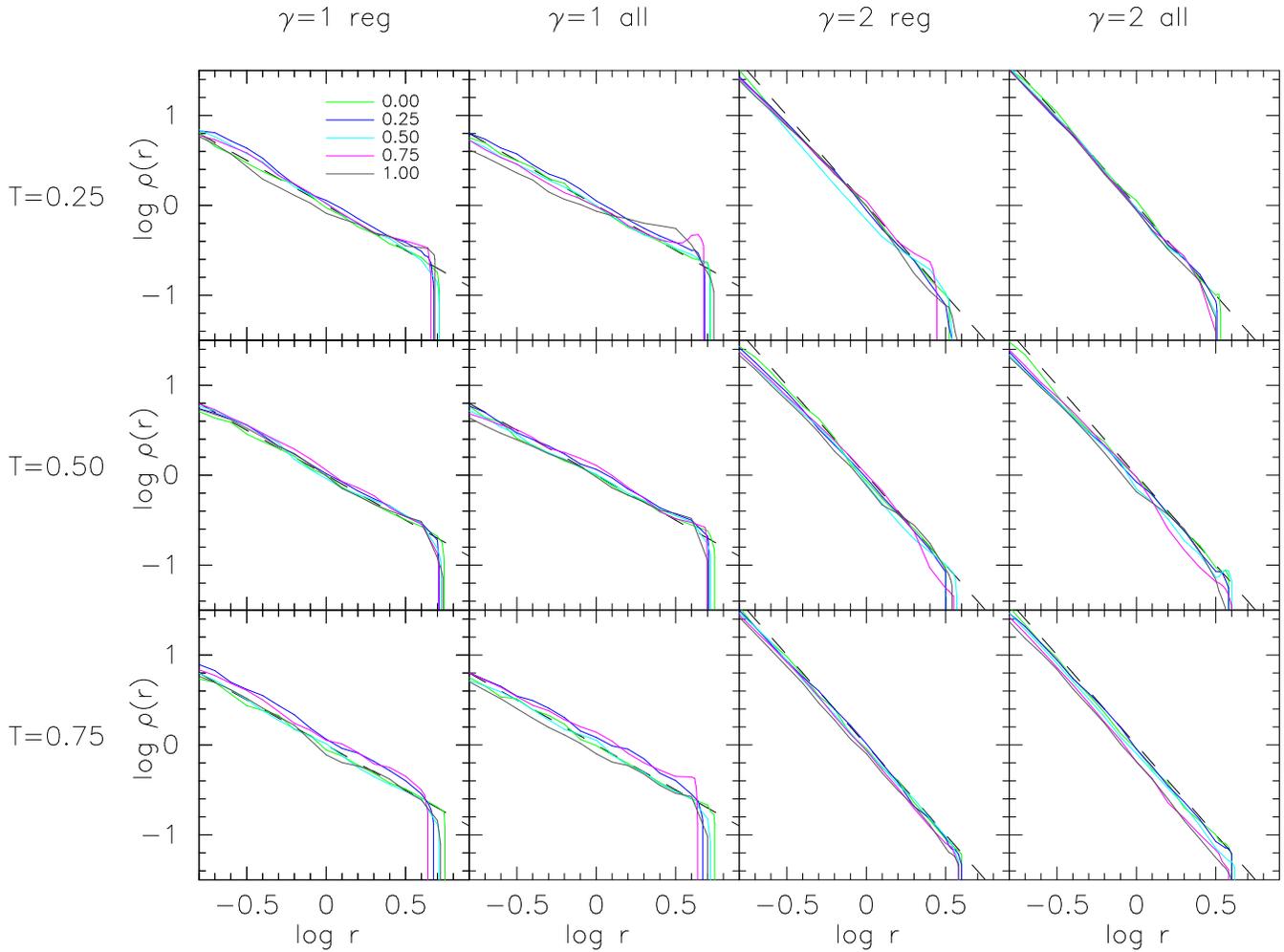}
\epsscale{1.0}
\caption{Evolution of density profiles for the orbital solutions
of Figure~\ref{MYP_Figure4}.
Different colored curves correspond to the times shown in the upper
left-hand panel, in units of the total integration time.
Dotted lines are the density profile of the underlying mass model.
\label{MYP_Figure6}}
\end{figure*}

As our primary index of evolution,
we computed the axis ratios of the models as a function of
radius using the iterative procedure described by Dubinski \&
Carlberg (1991), as follows. 
(i) The moment of inertia matrix of the particles enclosed 
by a sphere of radius $r$ is calculated.
(ii) Axis ratios are assigned as  $a = \sqrt{m_{11}/m_{max}}$, 
$b = \sqrt{m_{22}/m_{max}}$, $c=\sqrt{m_{33}/m_{max}}$,
where $m_{ii}$ are the principal moments of inertia and
$m_{max} = \mbox{max}\{ m_{11},~m_{22},~m_{33}\}$.
(iii) Particles are enclosed within the ellipsoid 
$x^2/a^2 + y^2/b^2 + z^2/c^2 = r^2$ and step (ii) is iterated,
until the axis ratios converge.

Figures~\ref{MYP_Figure5} and \ref{MYP_Figure6}
show the evolution of the axis ratios and density profiles 
of the various solutions.
None of the density profiles showed signficant evolution:
the radial distribution of mass did not evolve for any of the models.
However some of the solutions showed significant evolution in their 
shapes.
For $T=0.5$ (maximum triaxiality) and $\gamma=(1,2)$, 
we observed no significant evolution
in the axis ratios, either for the regular-orbit solutions or the 
solutions containing chaotic orbits.
We conclude that the maximally-triaxial solutions represent 
bona-fide equilibria.
For the weak-cusp model with $T=0.25$ (nearly oblate), 
there is some evolution at late times in the large-radius axis ratios,
while the nearly-oblate model with $\gamma=2$ hardly evolves.
Contour plots of the particle distribution suggest that the
evolution for $\gamma=1$ is due
to the relatively large errors in the cell masses at large radii;
as time goes on these errors propagate to smaller radii.
In spite of these fluctuations, however, 
we judged both of the $T=0.25$ solutions to be stable.

\begin{figure*}[t]
\epsscale{1.3}
\plotone{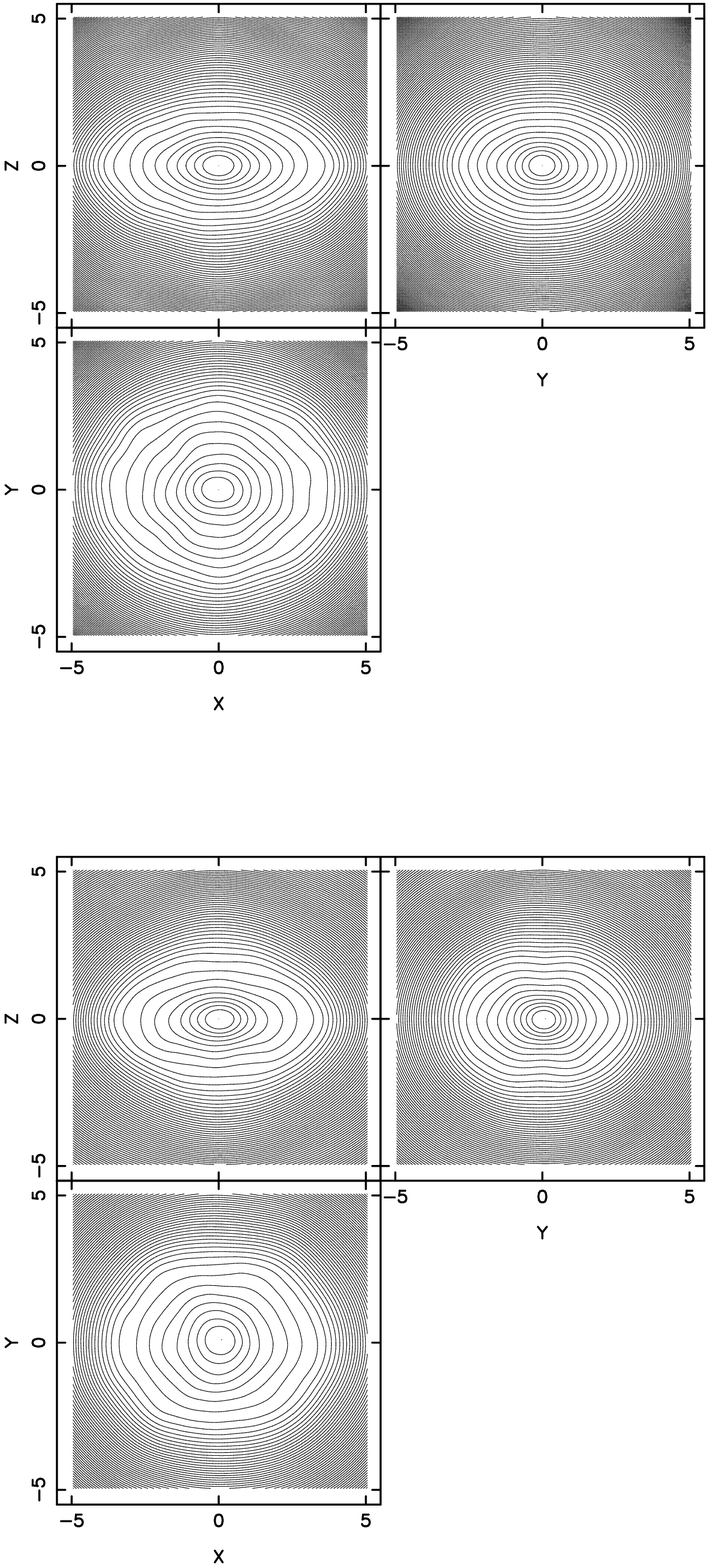}
\epsscale{1.3}
\caption{\label{MYP_Figure7}
Contours of the projected density of the solution with $\gamma=1$,
$T=0.50$ and only regular orbits. The upper panels show the contours
at $t=0$, and the lower panels show the contours at the end of the 
integration.
This model was judged stable based on the lack of significant
time evolution of the axis ratios (Figure 5); 
note the slight evolution of the contour shapes at the largest radii.}
\end{figure*}

By contrast, {\it all} solutions with $T=0.75$ (nearly prolate) were
found to exhibit substantial evolution, 
reaching nearly axisymmetric (oblate) shapes by the final time step. 
(We note that for the weak-cusp solution with $T=0.75$,
even the initial, intermediate-to-long axis ratio deviated 
noticeably from that of the assumed mass model.)
We were unable to find any solutions with $T=0.75$ that did not
evolve toward complete axisymmetry.
We conclude that these solutions do not represent bona-fide equilibria.

Figure~\ref{MYP_Figure7} shows contours of the projected density of the
weak cusp models with $T=0.50$ and only regular orbits. The contours 
remain approximately ellipsoidal until the end of the integration.

\section{A model with outer angular constraints relaxed} \label{outernuclei}
\begin{figure*}[t]
\epsscale{1.3}
\plotone{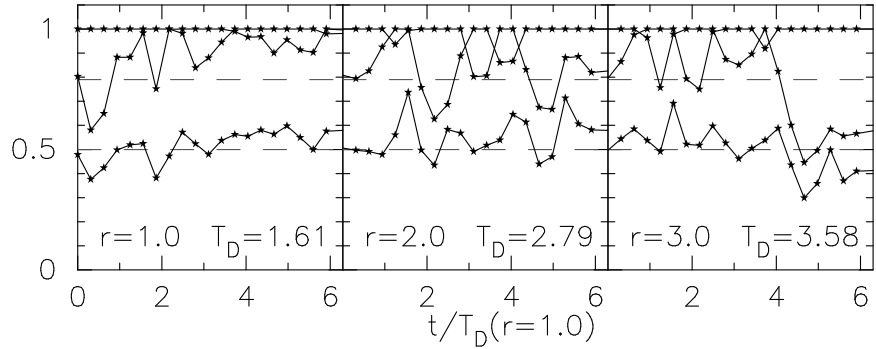}
\epsscale{1.3}
\caption{\label{MYP_Figure8}
Evolution of axis ratios of the weak cusp solution with the angular constraints
on the outer five shells relaxed.}
\end{figure*}

\begin{figure*}[t]
\epsscale{1.3}
\plotone{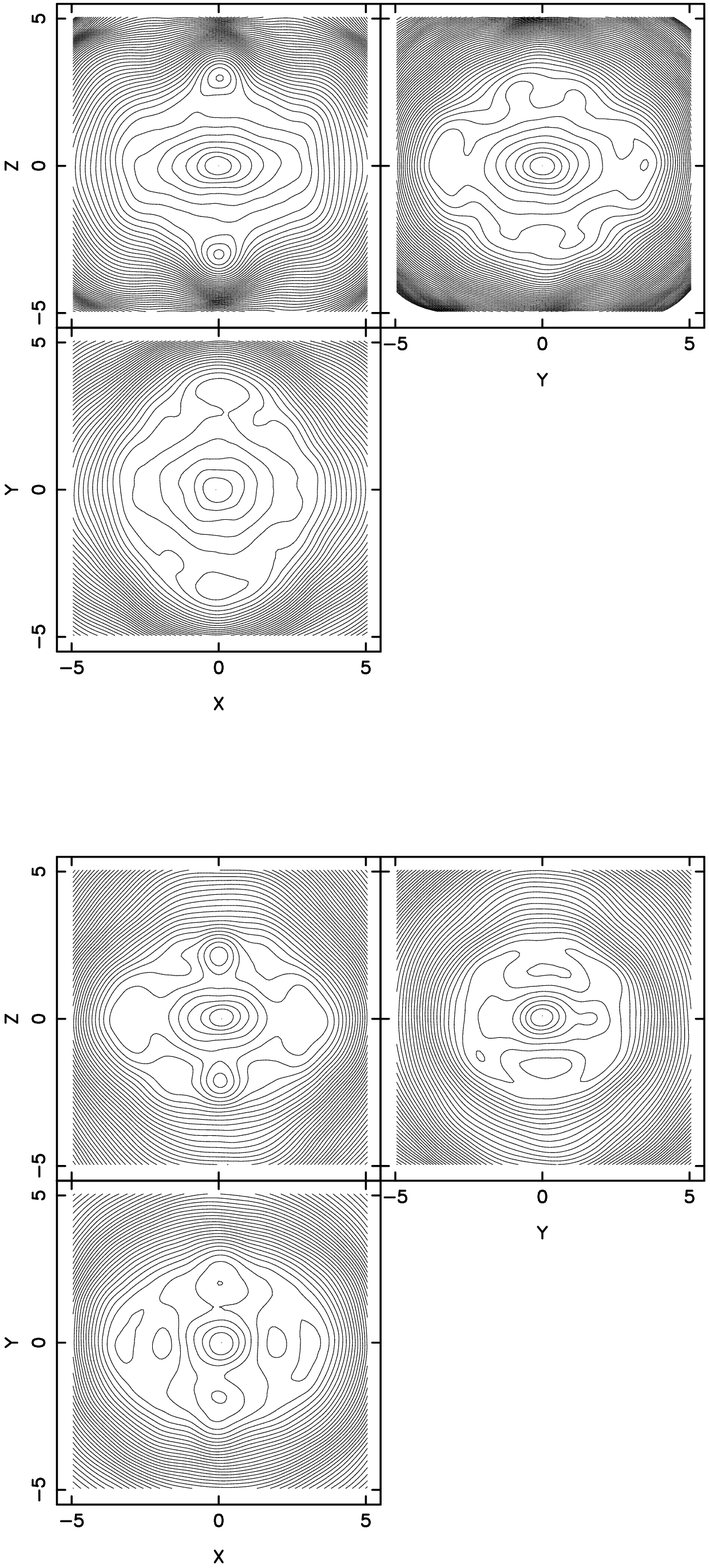}
\epsscale{1.3}
\caption{\label{MYP_Figure9}
Contours of the projected density of the weak cusp solution with
$T=0.50$ and only regular orbits.
The angular constraints of the 5 outer shells are ignored.
The upper panels show the contours
at $t=0$, and the lower panels show the contours at the end of the integration.}
\end{figure*}

We mentioned above that relaxing the angular constraints in the
outermost shells led to quadratic programming solutions with
very small errors in the innermost mass cells.
However these solutions generally exhibited large changes in their
shapes when evolved forward.
We illustrate this in the case of a weak cusp nucleus
with $T=0.50$ and only regular orbits.
In the orbital solution for this model,
we relaxed the angular constraints in the outer 5 shells, i.e.
only the masses of these shells were fit by the quadratic
programming routine, not the individual cell masses. 
By relaxing the outermost angular constraints, we were able
to find a solution in which $\log(\Delta) = -2.1$ for the inntermost
shells.
Figure~\ref{MYP_Figure8} shows the evolution of axis lengths of the
model.
The axis ratios show large fluctuations, especially at large radii.
Figure~\ref{MYP_Figure9} shows contours of the projected density. 
At $t=0$, the contours are ellipsoidal only at small radii;
the large-radius contours are irregular due to the large cell
mass discrepancies in the outer shells. 
During the integration, this model 
shows significant evolution at large radii, which in turn
affects the contours at small radii.
The final model looks very different from the initial model.
We found similar behavior in other orbital solutions when
the outermost constraints were relaxed.

\begin{figure*}[t]
\epsscale{1.4}
\plotone{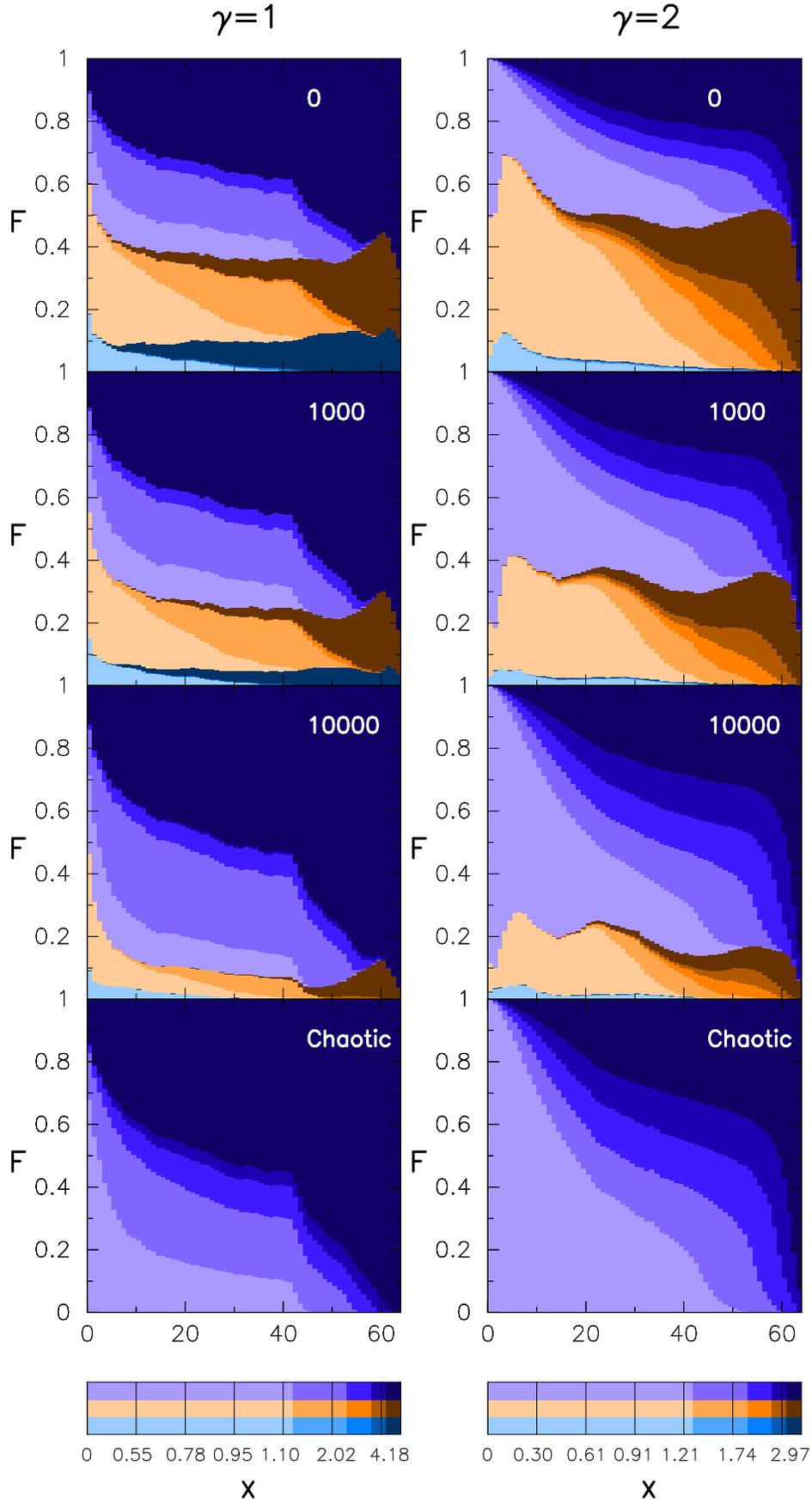}
\epsscale{1.4}
\caption{Cumulative mass fractions $F$ contributed by different kinds of orbits
to different shells of the triaxial solutions in which the contribution from
chaotic orbits has been maximized.
The value of $W_C$ is indicated in the upper right hand corner.
Left column: $T=0.5, \gamma=1$. 
Right column: $T=0.5, \gamma=2$.
\label{MYP_Figure10}}
\end{figure*}

\begin{figure*}[t]
\epsscale{2.0}
\plotone{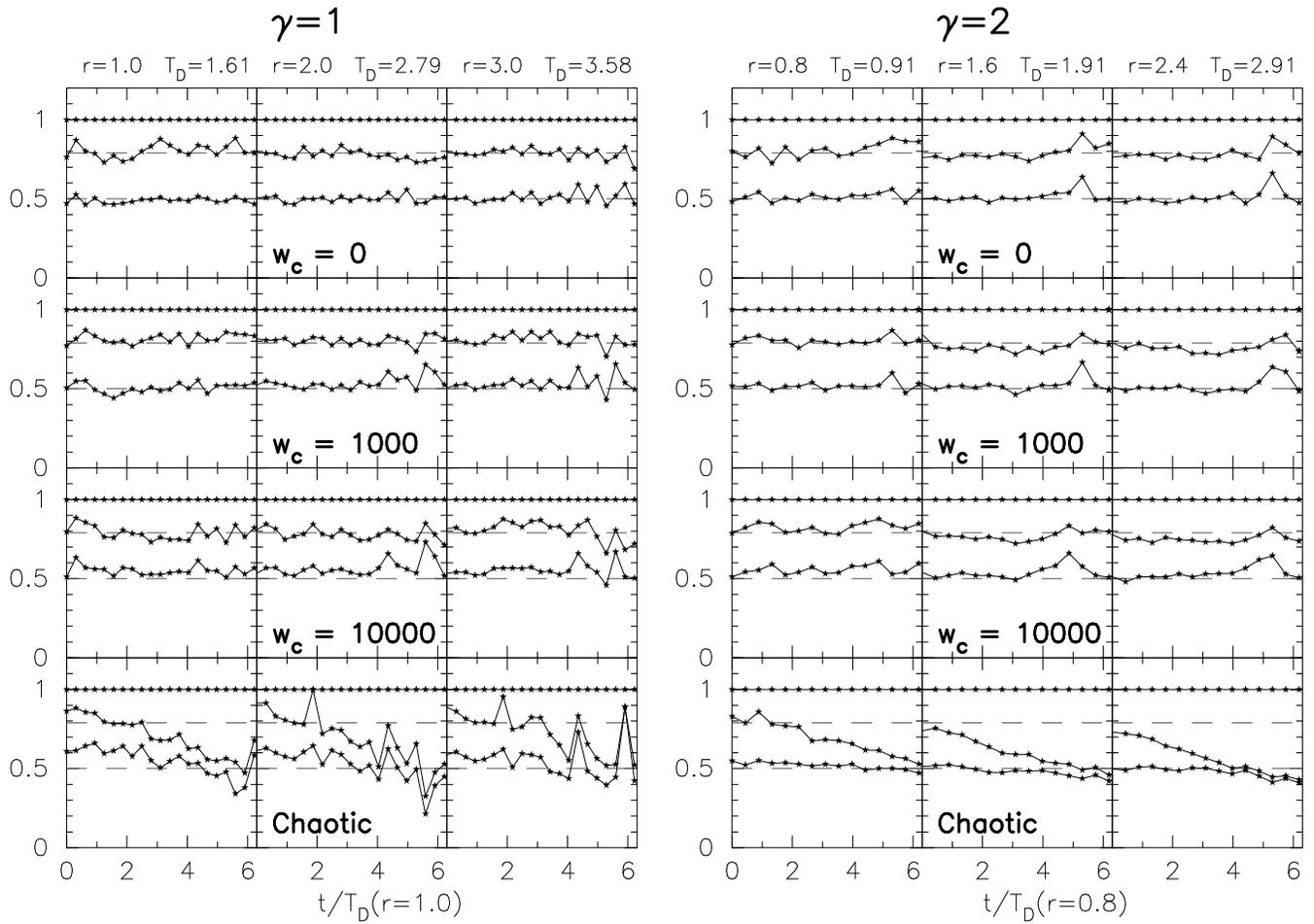}
\epsscale{1.0}
\caption{Evolution of the axis ratios of the orbital solutions
illustrated in Figure~\ref{MYP_Figure10}.
Columns on the left show models with $\gamma=1, T=0.50$ and various
values of $W_C$ at $r=1.0$ (first column), $r=2.0$ (second column), 
and $r=3.0$ (third column).
Columns on the right show models with $\gamma=2, T=0.50$.
\label{MYP_Figure11}}
\end{figure*}

\begin{figure*}[t]
\epsscale{1.3}
\plotone{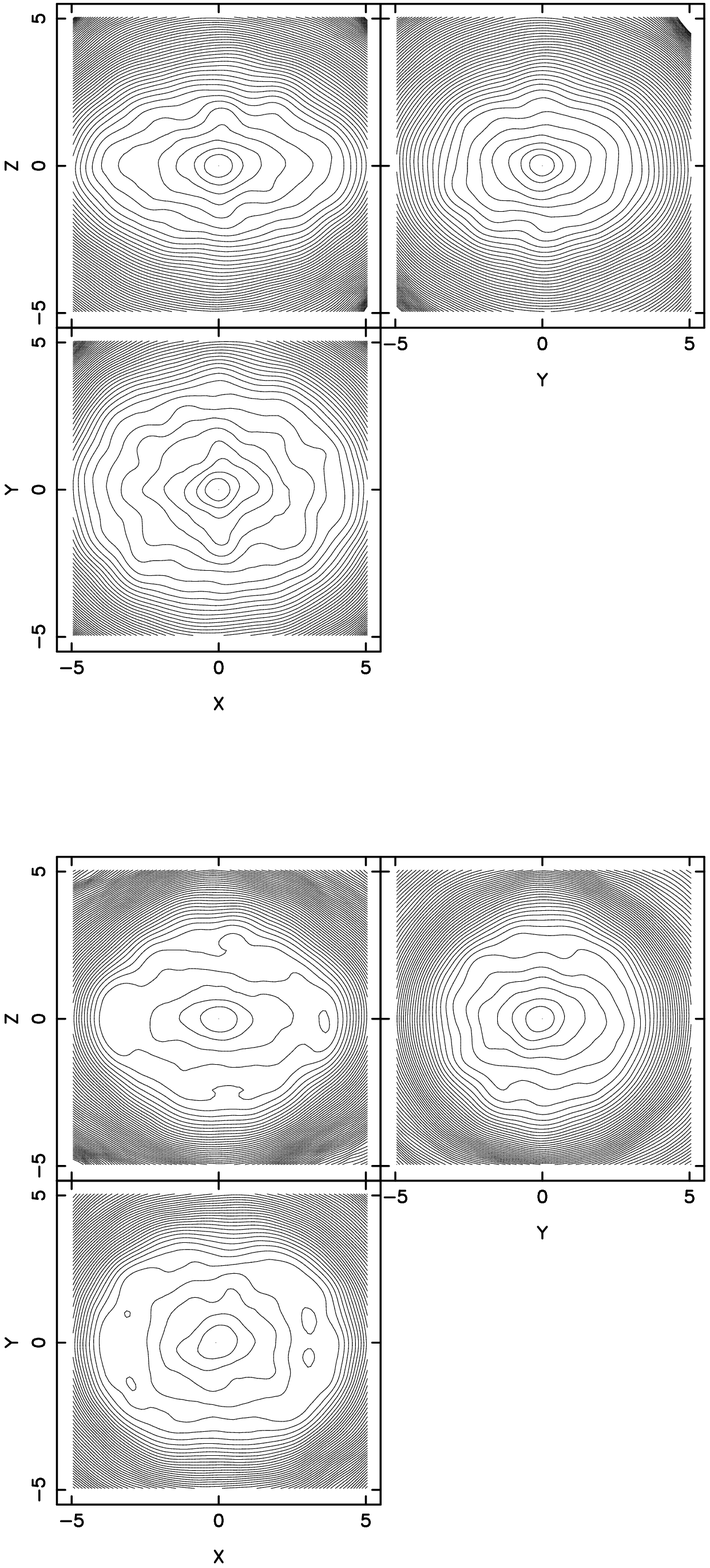}
\epsscale{1.3}
\caption{\label{MYP_Figure12}
Contours of the projected density of the weak cusp model with
$T=0.50$ and $W_C= 10000$.
The upper panels show the contours at $t=0$, and
the lower panels show the contours at the end of the integration.}
\end{figure*}

\section{Maximizing the contribution from Chaotic Orbits}
The existence of long-lived triaxial models containing
an abundance of chaotic orbits is particularly interesting:
stars on such orbits pass once per crossing time near
the center, greatly increasing the rate of interactions 
with the black hole compared with spherical or axisymmetric models.
We sought to maximize the chaotic content in our models
by minimizing the quantity:
\begin{eqnarray} 
\chi^2 = \sum_{l=1}^{N}(D_{l} - \sum_{i=1}^{M}B_{li}C_{i})^2 
	+ \sum_{i=1}^{M} W_i C_i 
\end{eqnarray}
instead of (\ref{rChi}).
Here $W_i$ is the ``penalty'' associated with the $i$th orbit;
increasing $W_i$ tends to decrease the contribution $C_i$ from
the $i$th orbit in the solution.
We set $W_i = 0 $ for the chaotic orbits and 
$W_i \equiv W_C = (0, 1000, 10000)$ for the regular orbits.
As $W_C$ increases, we expect the contribution from chaotic orbits
to increase, although possibly at the expense of the overall quality of
the fit to the cell masses.
We also constructed solutions containing only chaotic orbits.

Figure~\ref{MYP_Figure10} and Table~\ref{table7}
show the orbital content for weak and strong cusp
solutions with $T=0.50$ and different values of $W_C$. 
The bottom panel of Figure~\ref{MYP_Figure10} shows solutions 
containing only chaotic orbits.
As $W_C$ increases, the regular orbits are replaced
by high energy (low energy) chaotic orbits in the weak (strong) 
cusp solutions.
Evolution of the axis-ratios is shown in Figure~\ref{MYP_Figure11}.
Remarkably, only the models constructed from purely chaotic
orbits show substantial evolution in their axis ratios.
Inspection of the contour plots (e.g. Figure~\ref{MYP_Figure12})
does reveal some evolution away from elliptical isophotes
in the solutions with large $W_C$, but the triaxiality
appears robust. 
We conclude that chaotic mass fractions as large as $\sim 75\%$
or more might be consistent with long-lived triaxiality in galactic
nuclei.

\newpage

\begin{tablehere}
\caption{Orbital Content of Solutions with $T=0.5$ and Different $W_C$ 
\label{table7}}
\noindent\begin{tabular}{c|cccc}
\hline \hline
$$ & $z$-tubes & $x$-tubes & pyramids & chaotic\\ \hline 
$\gamma=1 : W_C = 0$ &  &  &  &\\
$r<0.5$ & 0.29 & 0.07 & 0.11 & 0.54\\ 
$r<1.0$ & 0.27 & 0.03 & 0.10 & 0.60\\
$r<1.5$ & 0.25 & 0.02 & 0.10 & 0.62\\ \hline
$\gamma=1 : W_C = 1000$ & &  &\\
$r<0.5$ & 0.24 & 0.06 & 0.09 & 0.61 \\ 
$r<1.0$ & 0.21 & 0.03 & 0.06 & 0.70\\
$r<1.5$ & 0.20 & 0.02 & 0.06 & 0.73\\ \hline
$\gamma=1 : W_C = 10000$ &  &  &  &\\
$r<0.5$ & 0.14 & 0.05 & 0.05 & 0.76\\ 
$r<1.0$ & 0.09 & 0.02 & 0.03 & 0.87\\
$r<1.5$ & 0.07 & 0.01 & 0.01 & 0.91\\ \hline

$\gamma=2 : W_C = 0$&  &  &\\ 
$r<0.4$ & 0.46& 0.05& 0.09& 0.40\\ 
$r<0.8$ & 0.45& 0.05& 0.06& 0.44\\
$r<1.2$ & 0.44& 0.05& 0.05& 0.46\\ \hline
$\gamma=2 : W_C = 1000$ & & & &\\
$r<0.4$ & 0.26 & 0.04 & 0.04 & 0.65 \\
$r<0.8$ & 0.30 & 0.03 & 0.03 & 0.64 \\
$r<1.2$ & 0.30 & 0.02 & 0.03 & 0.65 \\ \hline
$\gamma=2 : W_C = 10000$&  &  &\\ 
$r<0.4$ & 0.15& 0.03& 0.03& 0.79\\ 
$r<0.8$ & 0.18& 0.01& 0.02& 0.78\\
$r<1.2$ & 0.17& 0.01& 0.02& 0.79\\ \hline
\end{tabular}
\end{tablehere}

\section{Summary and Discussion}

We have shown that long-lived triaxial configurations are possible
for nuclei containing black holes.
Models with $T=0.5$ (maximally triaxial) and
$T=0.25$ (oblate/triaxial) were constructed and found to be stable,
retaining their non-axisymmetric shapes until the end
of the integration interval, equal to several
crossing times.
Models with $T=0.75$ (prolate/triaxial) were always
found to evolve rapidly to axisymmetry; we
speculate that prolate/triaxial nuclei do not
exist.
The evolution seen in the nearly-prolate models
does not appear to be a consequence
of orbital chaos; indeed, in our stable solutions,
we were able to replace a surprisingly large fraction
of the regular orbits by chaotic
orbits without inducing noticeable evolution in
their shapes.
We found that at least 50\%, and perhaps as
much as 75\%, of the mass could be placed on
chaotic orbits in the maximally triaxial and
oblate/triaxial solutions.
Such models violate Jeans's theorem in its standard
form (e.g. \cite{bt87}) but are consistent with a generalized
Jeans's theorem (\cite{mer99}) if we assume that the chaotic
building blocks are ``fully mixed,'' that is, that they approximate
a uniform population of the accessible phase space.
This appears to be the case for the chaotic orbits
in our models which have a very short mixing time.
While a sudden onset of chaos can effectively destroy triaxiality in
models containing a large population of regular box orbits
(\cite{meq98}; \cite{sel01}),
our work shows that at least the central parts of galaxies
containing black holes can remain triaxial even when dominated
by chaotic orbits.

Our results have possibly important implications for the 
rate at which stars are fed to supermassive black holes
in galactic nuclei.
In spherical or axisymmetric nuclei, the feeding rate
is determined by the rate at which stars on eccentric orbits
are scattered into the loss cone, the phase-space region
defined by orbits with pericenters lying within the
black hole's tidal disruption radius.
In the case of chaotic orbits in a triaxial nucleus,
each passage brings the star near to the center,
and the time required for a star to pass within
a distance $r_t$ of the black hole should scale
roughly as $r_t^{-1}$ (e.g. \cite{geb85}).
Thus even in the absence of gravitational scattering,
the loss cone would remain full and the feeding rate
could be orders of magnitude higher than in 
axisymmetric nuclei.
We will examine these ``chaotic loss cones'' 
in detail in an upcoming paper (Merritt \& Poon 2003).

While our results strengthen the case for triaxiality
in galactic nuclei,
the case for nuclear triaxiality could be made even 
more compelling by the detection of isophotal twists
or minor-axis rotation at the
very centers of galaxies.
Such observations will be challenging,
requiring two-dimensional data on an
angular scale that resolves the black
hole's sphere of influence.
Existing integral-field spectrographs
on ground-based telescopes (e.g. SAURON, 
\cite{bac01}) 
can only achieve this resolution for
the nearest galaxies.
Equally valuable would be $N$-body studies
demonstrating that triaxial nuclei
can form in realistic mergers.

M.Y. Poon would like to thank Andrew Mack for stimulating
discussions and constructive comments.
This work was supported by NSF grants AST 96-17088 and AST 00-71099
and by NASA grants NAG5-6037 and NAG5-9046. M.Y. Poon is grateful to 
the Croucher Foundation for a postdoctoral fellowship.

\bigskip\bigskip

\appendix
\section{Equivalent Spherical Models}
We define the equivalent spherical models to have mass density:
\begin{equation}
\rho_{\star}(r) = \left(\frac{r}{d}\right)^{-\gamma}, \ \ \ \ d^3=abc.
\end{equation}
The mass of the central black hole is set to one.
The potential is:
\begin{eqnarray}
\Phi(r) &=& 2\pi r - \frac{1}{r},\ \ \ \ \gamma=1 \label{appen1} \\
        &=& 4\pi d^2 \ln \left(\frac{r}{d}\right) - \frac{1}{r} - 4\pi d^2, \ \ \ \ \gamma=2. \label{appen2}
\end{eqnarray}
The constant terms in the expressions for the potential were obtained by taking 
the spherical limits of equation (3) of Paper I.
The isotropic distribution function $f(E)$ is given by
Eddington's formula,
\begin{eqnarray}
f(E) &=& \frac{\sqrt{2}}{4\pi^2} \int_{E}^{u} \frac{d\Phi}{\sqrt{\Phi-E}} \frac{d\rho}{d\Phi} \\
&=& \frac{\sqrt{2}}{4\pi^2}\left( \int^{u}_{E}\frac{d^2\rho}{d\Phi^2}\frac{\Phi}{\sqrt{\Phi-E}} + \lim_{\Phi \to u}  \frac{d\rho}{d\Phi} \frac{1}{\sqrt{\Phi-E}}\right)
\end{eqnarray}
and  $ u \equiv \lim_{r\to \infty}\Phi(r) $. 
We assume that the models extend to infinity.
In order to apply Eddington's formula, 
we need to express $\rho$ in terms of $\Phi$. 
For $\gamma=1$, we have:
\begin{eqnarray}
r(\Phi) = \frac{\Phi + \sqrt{\Phi^2 + 8d\pi}}{4d\pi}
\end{eqnarray}
and
\begin{eqnarray}
\rho(\Phi) = \frac{d}{r} = \frac{4d^2\pi}{\Phi + \sqrt{\Phi^2 + 8d\pi}}.
\end{eqnarray}
Thus
\begin{eqnarray}
\frac{d^2\rho}{d\Phi^2} &=& \frac{4d^2\pi}{(\Phi^2 + 8d\pi)^{\frac{3}{2}}}, \\
f(E) &=& \frac{\sqrt{2} d^2}{\pi} \int_{E}^{\infty} \frac{1}{(\Phi^2 + 8d\pi)^{\frac{3}{2}}
(\Phi - E)^{\frac{1}{2}}} d\Phi.
\end{eqnarray}
Similarly, for $\gamma=2$:
\begin{eqnarray}
r(\Phi) &=& \frac{1}{4\pi d^2} \frac{1}{\mbox{W}(u)}, \\
u &=& \frac{1}{4\pi d^2} \exp\left(-\frac{\Phi + 4\pi d^2 (1+\ln d)}{4\pi d^2}\right)
\end{eqnarray}
and
\begin{eqnarray}
\rho(\Phi) = \frac{d^2}{r^2} = 16 \pi^2 d^6 (\mbox{W}(u))^2.
\end{eqnarray}
$W(u)$ is Lambert's $W$ function; it is the inverse of the function
$u(W)=We^W$.
The distribution function is:
\begin{eqnarray}
f(E) = \frac{\sqrt{2}d^2}{d\pi^2}\int_{E}^{\infty} 
\frac{(\mbox{W}(u))^2(2 + \mbox{W}(u))}
{(1+\mbox{W}(u))^3 \sqrt{\Phi - E}} d\Phi.
\end{eqnarray}
For both models, the differential energy distribution is given by:
\begin{eqnarray}
M(E)dE &=& 16\pi^2p(E)f(E) dE,\\
p(E) &=& \int_0^{\Phi^{-1}(E)}\sqrt{2(E-\Phi(r)}\ dr.
\end{eqnarray}


\clearpage
\begin{thebibliography}{}

\bibitem[Bacon et al. 2001]{bac01}
	Bacon, R. et al. 2001,
	MNRAS, 326, 23

\bibitem[Binney \& Tremaine 1987]{bt87}
        Binney, J. \& Tremaine, S. 1987,
        Galactic Dynamics
        (Princeton: Princeton University Press)

\bibitem[Chandrasekhar 1969]{chandra69}
	Chandrasekhar, S. 1969,
	Ellipsoidal Figures of Equilibrium
	(Dover : New York)

\bibitem[Crane et al. 1993]{cra93}
        Crane, P. et al. 1993,
        AJ, 106, 1371

\bibitem[Dubinski \& Carlberg 1991]{duc91}
        Dubinski, J. \& Carlberg, R. 1991,
        ApJ, 378, 496

\bibitem[Erwin \& Sparke 2002]{ers02}
	Erwin, P. \& Sparke, L. S. 2002,
	AJ, 124, 65

\bibitem[Ferrarese et al. 1994]{fer94}
	Ferrarese, L., Van der Bosch, F.C., Ford, H.C., Jaffe, W., 
	\& O'Connell, R.W. 1994, 
	AJ, 108, 1598

\bibitem[Ferrarese et al. 2001]{fer01}
	Ferrarese, L., Pogge, R. W., Peterson, B. M., Merritt, D., 
	Wandel, A., \& Joseph, C. L. 2001,
	ApJ, 555, L79


\bibitem[Gerhard \& Binney 1985]{geb85}
	Gerhard, O. E. \& Binney, J. 1985,
	MNRAS, 216, 467

\bibitem[Hairer \& Wanner 1996]{hairer96}
	Hairer, E., \& Wanner, G. 1996,
	Solving Ordinary Differential Equations II.
	(Berlin : Springer)


\bibitem[Ho 1999]{ho99}
	Ho, Luis C., ``Supermassive Black Holes in Galactic Nuclei'',
	to appear in Observational Evidence for Black Holes in the
	Universe, ed. S.K. Chakraberti (Dordrecht : Kluwer)




\bibitem[McLure \& Dunlop 2002]{mcd02}
	McLure, R. J. \& Dunlop, J. S. 2002,
	MNRAS, 331, 795

\bibitem[Merritt 1999]{mer99}
	Merritt, D. 1999,
	PASP, 111, 129

\bibitem[Merritt 1980]{mer80}
	Merritt, D. 1980,
	ApJS, 43, 435

\bibitem[Merritt \& Ferrarese 2001]{mef01}
	Merritt, D. \& Ferrarese, L. 2001,
	in ``The Central Kiloparsec of Starbursts and AGN: 
	The La Palma Connection,'' ASP Conference Proceedings Vol. 249,
	ed. J. H. Knapen, J. E. Beckman, I. Shlosman, and T. J. Mahoney. 
	cisco: Astronomical Society of the Pacific) p. 335.,  

		

\bibitem[Merritt \& Fridman 1996]{mef96} 
        Merritt, D. \& Fridman, T. 1996, 
        ApJ, 460, 136

\bibitem[Merritt \& Poon 2003]{mep03}
	Merritt, D. \& Poon, M. 2003,
	in preparation.

\bibitem[Merritt \& Quinlan 1998]{meq98}
        Merritt, D. \& Quinlan, G. D. 1998,
        ApJ, 498, 625

\bibitem[Merritt \& Valluri 1996]{mev96}
	Merritt, D., \& Valluri, M. 1996,
	ApJ, 471, 82

\bibitem[Merritt \& Valluri 1999]{mev99}
        Merritt, D. \& Valluri, M. 1999,
        AJ, 118, 1177

\bibitem[Peterson 2002]{pet02}
	Peterson, B. M. 2002,
	astro-ph/0210639

\bibitem[Poon \& Merritt 2001]{pom01}
        Poon, M.Y. \& Merritt, D. 2001,
        ApJ, 549, 192 (Paper I)

\bibitem[Poon \& Merritt 2002]{pom02}
        Poon, M.Y. \& Merritt, D. 2002,
        ApJ Letter accepted (Paper II)

\bibitem[Sambhus \& Sridhar 2000]{sas00}
	Sambhus, N. \& Sridhar, S. 2000,
        ApJ, 542, 143

\bibitem[Schwarzschild 1979]{sch79}
        Schwarzchild, M. 1979, ApJ, 232, 236

\bibitem[Schwarzschild 1982]{sch82}
	Schwarzschild, M. 1982, ApJ, 263, 599

\bibitem[Schwarzschild 1993]{sch93} 
        Schwarzschild, M. 1993, ApJ, 409, 563

\bibitem[Sellwood 2001]{sel01}
	Sellwood, J. 2001,
	astro-ph/0107353

\bibitem[Smith \& Miller 1982]{sm82}
	Smith, B.F. \& Miller, R.H. 
	1982, ApJ, 257, 103
	
\bibitem[Springel, Yoshida \& White 2001]{syw01}
        Springel, V., Yoshida, N. \& White, S. D. M. 2001,
        NewA, 6, 79

\bibitem[Statler 1987]{stat87}
	Statler, T.S. 1987, 
	ApJ, 321, 113 



\end{thebibliography}
\end{document}